\documentclass[10pt]{article}
\usepackage{multicol}
\usepackage{graphicx}
\usepackage{amsmath}

\begin{document}

\begin{center}
{\Large\bf Event patterns extracted from anisotropic spectra of
charged particles produced in Pb-Pb collisions at 2.76 TeV}

\vskip0.75cm

Ya-Hui Chen and Fu-Hu Liu{\footnote{E-mail: fuhuliu@163.com;
fuhuliu@sxu.edu.cn}}

{\small\it Institute of Theoretical Physics \& State Key
Laboratory of Quantum Optics and Quantum Optics Devices,

Shanxi University, Taiyuan, Shanxi 030006, China}
\end{center}

\vskip0.5cm

{\bf Abstract:} Event patterns extracted from anisotropic spectra
of charged particles produced in lead-lead collisions at 2.76 TeV
are investigated. We use an inverse power-law resulted from the
QCD calculus to describe the transverse momentum spectrum in the
hard scattering process, and a revised Erlang distribution
resulted from a multisource thermal model to describe the
transverse momentum spectrum and anisotropic flow in the soft
excitation process. The pseudorapidity distribution is described
by a three-Gaussian function which is a revision of the Landau
hydrodynamic model. Thus, the event patterns at the kinetic
freeze-out are displayed by the scatter plots of the considered
particles in the three-dimensional velocity, momentum, and
rapidity spaces.
\\

\\


\vskip1.0cm

\begin{multicols}{2}

{\section{Introduction}}

Chemical and kinetic freeze-outs are two main stages of system
evolution in high energy collisions. In the stage of chemical
freeze-out, the ratios of different particle yields in the
interaction system are invariant, and the collisions between (or
among) these particles are inelastic. Meanwhile, the processes of
particle decay and production keep in a state of dynamic
equilibrium. In the stage of kinetic or thermal freeze-out, the
momentum and transverse momentum distributions of different
particles are invariant, and the collisions between (or among)
these particles are elastic. Meanwhile, the interaction system is
expected to stay in a state of thermal equilibrium.

Generally, two main stages of particle production are considered.
The hard scattering is believed not to undergo the chemical
freeze-out. The other stage, namely, the soft excitation process
is considered undergoing the chemical and kinetic freeze-outs. The
hard scattering process is actually described by the theory of
strong interactions, Quantum Chromodynamics (QCD), namely by its
perturbative calculations, in particular by the QCD calculus
[1--3] we use here. The soft excitation process is often described
by the thermal-related models [4--7] or hydrodynamic-related
models [8--19].

To understand the properties of chemical and kinetic freeze-outs
as well as soft process, one can use different models to describe
different experimental quantities such as the particle ratio,
transverse momentum spectrum, rapidity (or pseudorapidity)
spectrum, elliptic flow or higher order flow distribution,
dependence of elliptic flow on transverse momentum, and others.
According to these quantities, one can extract some parameters and
structure the event patterns at different conditions by using the
scatter plots of the considered particles. These event patterns
can give us relatively whole and objective pictures of the
interaction system at the freeze-outs and are helpful for us to
better understand these quantities. The dependences of event
patterns on the particle type, event centrality, and collision
energy are particularly important and interesting.

In our previous works [20--23], the event patterns extracted from
the transverse momentum and rapidity (or pseudorapidity) spectra
were studied, and an isotropic assumption in the transverse plane
was used. In this paper, a non-zero elliptic flow is considered in
the model treatment. We use a two-component model [1--3, 24] to
describe the transverse momentum spectrum and elliptic flow
measurements [25--31]. A three-Gaussian function being a revision
[17--19] of the Landau hydrodynamic model [32, 33], which results
in a Gaussian function [8--17], is used to describe the
pseudorapidity distribution. Based on the description of
experimental anisotropic spectra of charged particles produced in
lead-lead (Pb-Pb) collisions at the center-of-mass energy per
nucleon, $\sqrt{s_{NN}}$, of 2.76 TeV [25--31], event patterns are
structured in the three-dimensional velocity, momentum, and
rapidity spaces.
\\

{\section{The model and method}}

Before introducing the model and method, we have to structure a
coordinate system and definite some variables. Let the collision
point be the original $O$, one of the beam directions be the $Oz$
axis, and the reaction plane be the $xOz$ plane. We can structure
a right-handed coordinate system in which the $Ox$ axis is along
the impact parameter, $Oy$ axis is perpendicular to the $xOz$
plane, and the transverse plane is the $xOy$ plane. Let $\beta_x$,
$\beta_y$, and $\beta_z$ ($p_x$, $p_y$, and $p_z$) denote the
velocity (momentum) components on the $Ox$, $Oy$, and $Oz$ axes,
respectively; $Y_1$, $Y_2$, and $Y$ denote the rapidities defined
due to energy $E$ and $p_x$, $E$ and $p_y$, as well as $E$ and
$p_z$, respectively. To obtain the event patterns in the
three-dimensional velocity ($\beta_x-\beta_y-\beta_z$), momentum
($p_x-p_y-p_z$), and rapidity ($Y_1-Y_2-Y$) spaces at a given
condition, we need at least the transverse momentum $p_T$ and
rapidity $Y$ (or pseudorapidity $\eta$) spectra to extract the
values of related parameters.

It should be noted that the elliptic flow $v_2$ is not zero in
most cases such as in non-central collisions, which renders
anisotropic flow in the $xOy$ plane in the rest source frame. That
is, the assumption of isotropic emission in the $xOy$ plane used
in our previous works [20--23] is only an approximate treatment
and should be revised due to non-zero anisotropic flow. To include
anisotropic flow in the event patterns, we have to use
simultaneously $p_T$, $v_2$, and $Y$ (or $\eta$) spectra for
extracting parameters and structuring event patterns. Generally,
the effect of isotropic flow such as the radial flow is already
included in the $p_T$ spectrum and does not need to be considered
particularly.

Generally, different models use different ideas and methods to
treat the same collisions. Some of them are even inconsistent each
other. The model used in the present work is a hybrid model which
consists of the QCD calculus [1--3] for a wide $p_T$ spectrum
contributed by the hard scattering process, a multisource thermal
model [24] for a narrow $p_T$ spectrum contributed by the soft
excitation process, and a revised Landau hydrodynamic model
[17--19] for $Y$ or $\eta$ spectrum. As believed, there are
generally two main processes, the hard scattering process and the
soft excitation process, in particle productions in high energy
collisions.

The hard scattering process happens in the early stage and between
two valence quarks. Particles produced in the hard process
distribute in a wide $p_T$ range. According to the QCD calculus
[1--3], the hard process contributes an inverse power-law. One has
the following function form
\begin{equation}
f_1(p_T)=Ap_T \bigg(1+\frac{p_T}{p_0} \bigg)^{-n},
\end{equation}
where $A$ denotes the normalization constant which depends on the
free parameters $p_0$ and $n$. Because of the limitation of
normalization, one has naturally $\int_0^{\infty} f_1(p_T)
dp_T=1$.

The soft excitation process happens in the later stage and between
two (or among three or more) gluons and/or sea quarks. Particles
produced in the soft process distribute in a narrow $p_T$ range
(up to 2--3 GeV/$c$). According to the multisource thermal model
[24], we can use the Erlang distribution
\begin{equation}
f_2(p_{T})=\frac{p_T^{m-1}}{(m-1)!\langle p_{Ti} \rangle^m} \exp
\bigg(- \frac{p_{T}}{\langle p_{Ti} \rangle} \bigg)
\end{equation}
to describe the $p_T$ spectrum contributed by the soft process,
where $\langle p_{Ti} \rangle$ and $m$ are free parameters which
describe the mean contribution of each source (partons) and the
number of sources (partons) respectively. Naturally,
$\int_0^{\infty} f_2(p_T) dp_T=1$.

Analytically, we can use a superposition of the inverse power-law
and the Erlang distribution to describe the $p_T$ spectra of
final-state particles. In fact, we have the normalized
distribution
\begin{equation}
f_0(p_{T})=k_1f_{1}(p_{T})+(1-k_1)f_{2}(p_{T}),
\end{equation}
where $k_1$ denotes the contribution ratio of the inverse
power-law, and $1-k_1$ denotes naturally the contribution ratio of
the Erlang distribution. Obviously, this superposition obeys
$\int_0^{\infty} f_0(p_T) dp_T=1$. To give a comparison with the
non-normalized experimental data, the normalized constant
($N_{p_T}$) is needed.

In the Monte Carlo method, we can obtain $p_T$ due to the above
functions $f_1(p_T)$ and $f_2(p_T)$. Let $R$, $R_0$, and $r_i$
denote the random numbers distributed evenly in $[0,1]$. The
values of $p_T$ in Eq. (1) or in the first component in Eq. (3)
can be obtained by
\begin{equation}
\int_0^{p_{T}}f_{1}(p_{T})dp_{T} <R
<\int_0^{p_{T}+dp_{T}}f_{1}(p_{T}) dp_{T}.
\end{equation}
The values of $p_T$ in Eq. (2) or in the second component in Eq.
(3) can be obtained by
\begin{equation}
p_T=-\langle p_{Ti} \rangle \sum_{i=1}^{m} \ln r_i = -\langle
p_{Ti} \rangle \ln \prod_{i=1}^m r_i.
\end{equation}
Then, $f_1(p_T)$, $f_2(p_T)$, and $f_0(p_T)$ can be obtained by
statistics.

Under the assumption of isotropic emission, we have the momentum
components in the $xOy$ plane to be
\begin{equation}
p_x=p_T \cos \varphi=p_T \cos(2\pi R_0),
\end{equation}
\begin{equation}
p_y=p_T\sin \varphi =p_T \sin(2\pi R_0),
\end{equation}
where
\begin{equation}
\varphi=\arctan \bigg (\frac{p_y}{p_x} \bigg)= 2\pi R_0
\end{equation}
denotes the azimuthal angle. The isotropic emission results in the
elliptic flow
\begin{equation}
v_2=\langle \cos(2\varphi) \rangle =\bigg\langle
\frac{p_x^2-p_y^2}{p_x^2+p_y^2} \bigg\rangle =0,
\end{equation}
where $\langle ... \rangle$ denotes the averaging events.

However, the experimental $v_2$ in most cases is not equal to zero
[26, 28, 34--37]. This means that, in the $xOy$ plane, we have to
consider an anisotropic emission. The anisotropic emission
enlightens us to consider the interactions between (or among) the
isotropic sources. These interactions result in the deformation
and movement of the isotropic source. The deformation means
expansion and compression, and the movement can be along the
positive or negative axis direction.

Considering the deformation and movement of the source, $p_x$ and
$p_y$ obtained above are revised to
\begin{equation}
P_x = a_xp_x+b_x,
\end{equation}
\begin{equation}
P_y = a_yp_y+b_y,
\end{equation}
where $a_x$ ($a_y$) and $b_x$ ($b_y$) denote the deformation and
movement of the source in the $Ox$ ($Oy$) axis direction
respectively. The two equations are ascribed to the revised Erlang
distribution. They reflect approximately the mean effect of
interactions between (or among) the sources. The interactions are
described by $a_{x,y}$ and $b_{x,y}$ which are very small and can
be concretely seen in the next section. Generally, $a_{x,y}>1$
($<1$) means an expansion (compression), $b_{x,y}>0$ ($<0$) means
a movement along the positive (negative) axis direction. The
introduction of $a_{x,y}$ and $b_{x,y}$ results in a revised
Erlang distribution which can be obtained by the Monte Carlo
method.

Because of only the relative deformation of the source being
interested, we can require the minimum in $a_x$ and $a_y$ to be
fixed to 1, and the other one to be equal to or greater than 1.
Meanwhile, some sources can move along the positive axis
direction, and others can move along the negative axis direction.
Due to the revision on the momentum components, we have a new
expression for $v_2$,
\begin{equation}
v_2=\bigg\langle \frac{P_x^2-P_y^2}{P_x^2+P_y^2} \bigg\rangle .
\end{equation}
The transverse momentum after the transformation is
\begin{equation}
p_T=\sqrt{P_x^2+P_y^2},
\end{equation}
where the same symbol $p_T$ is used as that before the
transformation. Similarly, we use the symbol $p_{x,y}$ instead of
$P_{x,y}$ in the following discussions even for those after the
transformation. By using $v_2$ and $p_T$, we can study the
dependence of $v_2$ on $p_T$.

In the Landau hydrodynamic model and its revisions [8--19], the
$Y$ or $\eta$ distribution contributed by a given source can be
parameterized to a Gaussian function [8--17]. In the case of
considering $\eta$ distribution, we have
\begin{equation}
f_{\eta}(\eta)=\frac{1}{\sqrt{2\pi} \sigma_{\eta}} \exp \bigg[-
\frac{(\eta-\eta_C)^2}{2\sigma_{\eta}^2} \bigg],
\end{equation}
where $\eta_C$ denotes the mid-pseudorapidity or peak position and
$\sigma_{\eta}$ denotes the distribution width. If one regards a
Gaussian function for $Y$ or $\eta$ distribution as a result of
the Landau hydrodynamic model, any application of two or more
Gaussian functions is a revision of the model. This revision may
be caused by the leading particles or resonance production.

In refs. [17, 38, 39], one uses two or three Gaussian functions to
describe the $Y$ or $\eta$ distribution. In the case of using
three-Gaussian function which is the case in the present work, one
can write
\begin{align}
f_{\eta}(\eta) &= \frac{k_2}{\sqrt{2\pi} \sigma_{\eta_1}} \exp
\bigg\{-\frac{\big[\eta-(-\delta\eta)\big]^2}{2\sigma_{\eta_1}^2} \bigg\} \nonumber\\
&+ \frac{1-2k_2}{\sqrt{2\pi} \sigma_{\eta_2}} \exp \bigg[-
\frac{(\eta-\eta_C)^2}{2\sigma_{\eta_2}^2} \bigg] \nonumber\\
&+ \frac{k_2}{\sqrt{2\pi} \sigma_{\eta_1}} \exp \bigg[-
\frac{(\eta-\delta\eta)^2}{2\sigma_{\eta_1}^2} \bigg]
\end{align}
for symmetric collisions, where $\sigma_{\eta1}$ and
$\sigma_{\eta2}$ denote the distribution widths contributed by the
backward (or forward) source and central source respectively,
$k_2$ is the contribution ratio of the backward (or forward)
source, $1-2k_2$ is the contribution ratio of the central source,
and $-\delta \eta$ (or $\delta \eta$) is the pseudorapidity shift
of the peak position for the backward (or forward) source. When
comparing with experimental data, the normalization constant
($N_{\eta}$) is needed. It should be noted that the sources
discussed in the $Y$ or $\eta$ spectra are larger than those
discussed in the $p_T$ or $v_2$ spectra. The two types of sources
are different. It should be noted that, actually, the use of two
or three Gaussian functions is somehow not a necessity, see e.g.
[8--16].

In the Monte Carlo method, let $R_{1-6}$ denote the random numbers
distributed evenly in $[0,1]$. As for the variable $\eta$ in the
first (backward), second (central), and third (forward) components
(sources) in the three-Gaussian function, we have
\begin{equation}
\eta=\sigma_{\eta1} \sqrt{-2\ln R_1} \cos(2\pi R_2) -\delta\eta,
\end{equation}
\begin{equation}
\eta=\sigma_{\eta2} \sqrt{-2\ln R_3} \cos(2\pi R_4) +\eta_C,
\end{equation}
and
\begin{equation}
\eta=\sigma_{\eta1} \sqrt{-2\ln R_5} \cos(2\pi R_6) +\delta\eta,
\end{equation}
respectively. The contribution ratios are determined by $k_2$,
$1-2k_2$, and $k_2$ for the first, second, and third components,
respectively.

On the conversion from $\eta$ to $Y$ distributions, we perform the
Monte Carlo method. According to the definition of $\eta$, i.e.
\begin{equation}
\eta=-\ln\tan \bigg(\frac{\vartheta}{2}\bigg),
\end{equation}
where $\vartheta$ denotes the polar angle, we have
\begin{equation}
\vartheta=2\arctan(e^{-\eta}).
\end{equation}
Then, the momentum component on the $Oz$ axis is
\begin{equation}
p_z= p_T \cot \vartheta.
\end{equation}
The energy is
\begin{equation}
E=\sqrt{p_T^2+p_z^2+m_0^2},
\end{equation}
where $m_0$ is the rest mass. In the case of considering charged
particles, $m_0$ is the average weighted by rest masses and yields
of different types of particles.

Further, the velocity components are
\begin{equation}
\beta_{x}=\frac{p_{x}}{E},
\end{equation}
\begin{equation}
\beta_{y}=\frac{p_{y}}{E},
\end{equation}
and
\begin{equation}
\beta_{z}=\frac{p_{z}}{E}.
\end{equation}
The rapidity
\begin{equation}
Y=\frac{1}{2}\ln \bigg( \frac{E+p_z}{E-p_z} \bigg).
\end{equation}
Similar to the rapidity defined by $E$ and $p_z$, we define the
rapidity
\begin{equation}
Y_1=\frac{1}{2}\ln \bigg( \frac{E+p_{x}}{E-p_{x}} \bigg)
\end{equation}
due to $E$ and $p_x$, and the rapidity
\begin{equation}
Y_2=\frac{1}{2}\ln \bigg( \frac{E+p_{y}}{E-p_{y}} \bigg)
\end{equation}
due to $E$ and $p_y$.

In the concrete calculation, we need firstly to fit the $p_T$
distribution, dependence of $v_2$ on $p_T$, and $\eta$ (or $Y$)
distribution to get the values of free parameters and
normalization constants. Then, we can use the values of free
parameters obtained by the fit in the first step to get the
discrete values of different kinds of kinematic variables. After
repeating 1000 times calculation for each case (centrality), we
can get the event patterns in three-dimensional
$\beta_x-\beta_y-\beta_z$, $p_x-p_y-p_z$, and $Y_1-Y_2-Y$ spaces.

It should be noted that the parameterizations for the $p_T$ and
$\eta$ (or $Y$) are independent of models, though three different
models are used in the parameterizations. The parameterizations
performed by us are only for the extraction of discrete values for
$p_T$ and $\eta$ (or $Y$). Although we can use the experimental
discrete values themselves, the parameterizations can extract more
discrete values in wider $p_T$ and $\eta$ (or $Y$) ranges in the
case of using the limited experimental ranges.
\\

{\section{Results and discussion}}

Figure 1 shows the $p_T$ dependence of the double-differential
spectra, $(1/N_{EV}) \cdot 1/(2\pi p_T) \cdot d^2N_{ch}/(d\eta
dp_T)$, of charged particles produced in Pb-Pb collisions at
$\sqrt{s_{NN}}=2.76$ TeV, where $N_{EV}$ and $N_{ch}$ denote the
numbers of events and charged particles respectively. The symbols
represent the experimental data of the ALICE Collaboration [25]
measured in the mid-pseudorapidity range, $|\eta|<0.8$, with nine
centrality classes, 0--5\%, 5--10\%, 10--20\%, 20--30\%, 30--40\%,
40--50\%, 50--60\%, 60--70\%, and 70--80\%. The different
centrality classes are scaled by different amounts marked in the
panel for the purpose of clarity. The solid curves for the
intermediate four cases (10--20\%, 20--30\%, 30--40\%, and
40--50\%) are our model results calculated by using the
superposition of the inverse power-law and revised Erlang
distribution due to Fig. 2 which will be discussed later. The
dashed curves for the nine centrality classes are our model
results calculated by using the superposition of the inverse
power-law and (unrevised) Erlang distribution. In the calculation
for charged particles, we take $m_0=0.174$ GeV/$c^2$ which is the
average rest mass obtained by us for weighting rest masses and
yields of different types of particles as given in [40]. The
values of free parameters ($p_{0}$, $n$, $k_1$, $m$, and $\langle
p_{Ti} \rangle$) and the normalization constant ($N_{p_T}$) for
both the solid and dashed curves, $\chi^2$ and degree of freedom
(dof$|_1$) in terms of $\chi^2$/dof$|_1$ for the solid curves, as
well as $\chi^2$ and dof$|_2$ in terms of $\chi^2$/dof$|_2$ for
the dashed curves are listed in Table 1. The values of $a_{x,y}$
and $b_{x,y}$ for the solid curves are listed in Table 2, which
are those used in Fig. 2. For the dashed curves, we have
$a_{x,y}=1$ and $b_{x,y}=0$. In particular, for the intermediate
four cases, the solid curves are possibly not the best fitted
results due to the constraint of Fig. 2 which determines $a_{x,y}$
and $b_{x,y}$, and the dashed curves are not the best fitted
results due to the constraint of the solid curves in which we use
simply $a_{x,y}=1$ and $b_{x,y}=0$ to obtain the dashed curves.
The solid and dashed curves are very similar to each other in most
cases. One can see that the model results describe approximately
the ALICE experimental $p_T$ spectra of charged particles measured
in different centrality classes in Pb-Pb collisions at
$\sqrt{s_{NN}}=2.76$ TeV. The effect of anisotropic flow on the
$p_T$ spectra is small and can be neglected in fact. The
properties of parameters will be discussed later.

The dependences of elliptic flow $v_2\{4\}$ on $p_T$ for charged
particles produced in Pb-Pb collisions at $\sqrt{s_{NN}}=2.76$ TeV
in $|\eta|<0.8$ for four centrality classes, 10--20\%, 20--30\%,
30--40\%, and 40--50\% are presented in Fig. 2, where $v_2\{4\}$
denotes the elliptic flow obtained by a specially appointed method
[26]. Generally, different methods give different $v_2$ with small
differences. The symbols represent the experimental data of the
ALICE Collaboration [26], and the curves are our model results
which are resulted from the inverse power-paw which contributes
zero $v_2$ and the revised Erlang distribution which show
anisotropic flow. The parameters for the inverse power-law do not
affect the results due to its zero $v_2$. The parameters ($1-k_1$,
$m$, $\langle p_{Ti} \rangle$, $a_{x,y}$, and $b_{x,y}$),
$\chi^2$, and dof for the revised Erlang distribution are obtained
and listed in Table 1 or 2. One can see that the model results
describe approximately the ALICE experimental data of the
dependence of $v_2$ on $p_T$ for charged particles in different
centrality classes in Pb-Pb collisions at $\sqrt{s_{NN}}=2.76$
TeV. The effect of $v_2$ is obvious, though its influence on the
$p_T$ spectrum is small. We shall discuss the properties of
parameters later.

The $p_T$ dependence of the double-differential spectra of
positively and negatively charged pions ($\pi^++\pi^-$), kaons
($K^++K^-$), and protons plus antiprotons ($p+\bar p$) produced in
Pb-Pb collisions at $\sqrt{s_{NN}}=2.76$ TeV in $|\eta|<0.8$ for
eight centrality classes, 0--5\%, 5--10\%, 10--20\%, 20--30\%,
30--40\%, 40--50\%, 40--60\%, and 60--80\% are displayed in Figs.
3(a), 3(b), and 3(c), respectively. The different centrality
classes are scaled by different amounts marked in the panel for
the purpose of clarity. The symbols represent the experimental
data of the ALICE Collaboration [27] and the solid curves for the
first seven cases are our model results due to Fig. 4 which will
be discussed later. For comparison, the model results
corresponding to the superposition of the inverse power-law and
(unrevised) Erlang distribution are displayed by the dashed
curves. The values of $p_{0}$, $n$, $k_1$, $m$, $\langle p_{Ti}
\rangle$, and $N_{p_T}$ for both the solid and dashed curves,
$\chi^2$ and dof$|_1$ for the solid curves, as well as $\chi^2$
and dof$|_2$ for the dashed curves are listed in Table 1. The
values of $a_{x,y}$ and $b_{x,y}$ for the solid curves are listed
in Table 2, which are the same as Fig. 4. For the first seven
cases, the solid curves are not the best fitted results due to the
constraint of Fig. 4 which determines $a_{x,y}$ and $b_{x,y}$, and
the dashed curves are not the best fitted results due to the
constraint of the solid curves in which we use simply $a_{x,y}=1$
and $b_{x,y}=0$ to obtain the dashed curves. The sole exception is
the dashed curves for the centrality 60--80\% in which there is no
particular constraint. One can see again that the model results
describe approximately the ALICE experimental $p_T$ spectra of
identified particles measured in different centrality classes in
Pb-Pb collisions at $\sqrt{s_{NN}}=2.76$ TeV. The effect of
anisotropic flow on the $p_T$ spectrum is small and can be
neglected in most cases. The parameters for the identified
particle spectra are a decomposition of the parameters for the
charged particle spectra. Because of the main component in charged
particles being $\pi^++\pi^-$, the parameters for the charged
particle spectra are determined by those for $\pi^++\pi^-$
spectra.

The dependences of elliptic flow{\footnote{The ``SP" notation
refers to the Scalar Product method [28] representing a
two-particle correlation technique which uses a gap of $|\Delta
\eta|>0.9$ between the identified hadron under consideration and
the reference particles.}} $v_2\{SP,|\Delta \eta|>0.9\}$ on $p_T$
for identified particles [(a) $\pi^++\pi^-$, (b) $K^++K^-$, and
(c) $p+\bar p$] produced in Pb-Pb collisions at
$\sqrt{s_{NN}}=2.76$ TeV in $|\eta|<0.8$ for seven centrality
classes, 0--5\%, 5--10\%, 10--20\%, 20--30\%, 30--40\%, 40--50\%,
and 50--60\% are presented in Fig. 4, where $v_2\{SP,|\Delta
\eta|>0.9\}$ denotes the elliptic flow obtained by another
specially appointed method [28]. The symbols represent the
experimental data of the ALICE Collaboration [28], and the curves
are our model results which contains the revised Erlang
distribution. The parameters for the inverse power-law do not
affect the results due to its zero $v_2$. The parameters with
$\chi^2$ and dof for the revised Erlang distribution can be found
in Tables 1 and 2 respectively. One can see again that the model
results describe approximately the ALICE experimental data of the
dependence of $v_2$ on $p_T$ for identified particles in different
centrality classes in Pb-Pb collisions at $\sqrt{s_{NN}}=2.76$
TeV. The effect of $v_2$ is obvious, though its influence on the
$p_T$ spectrum is small. The parameters for $\pi^++\pi^-$ spectra
determine those for charged particles spectra. We would like to
point out that the disagreement with the data on the right side of
the $v_2(p_T)$ distribution is caused by the lack for revising the
inverse power-law. It is expected that an improvement can be
reached if we revise the inverse power-law as what we do for the
Erlang distribution. However, in this case, four more parameters
will be added to the fit procedure, which complicates the study.

\begin{figure*}
\vskip.0cm \begin{center}
\includegraphics[width=15.cm]{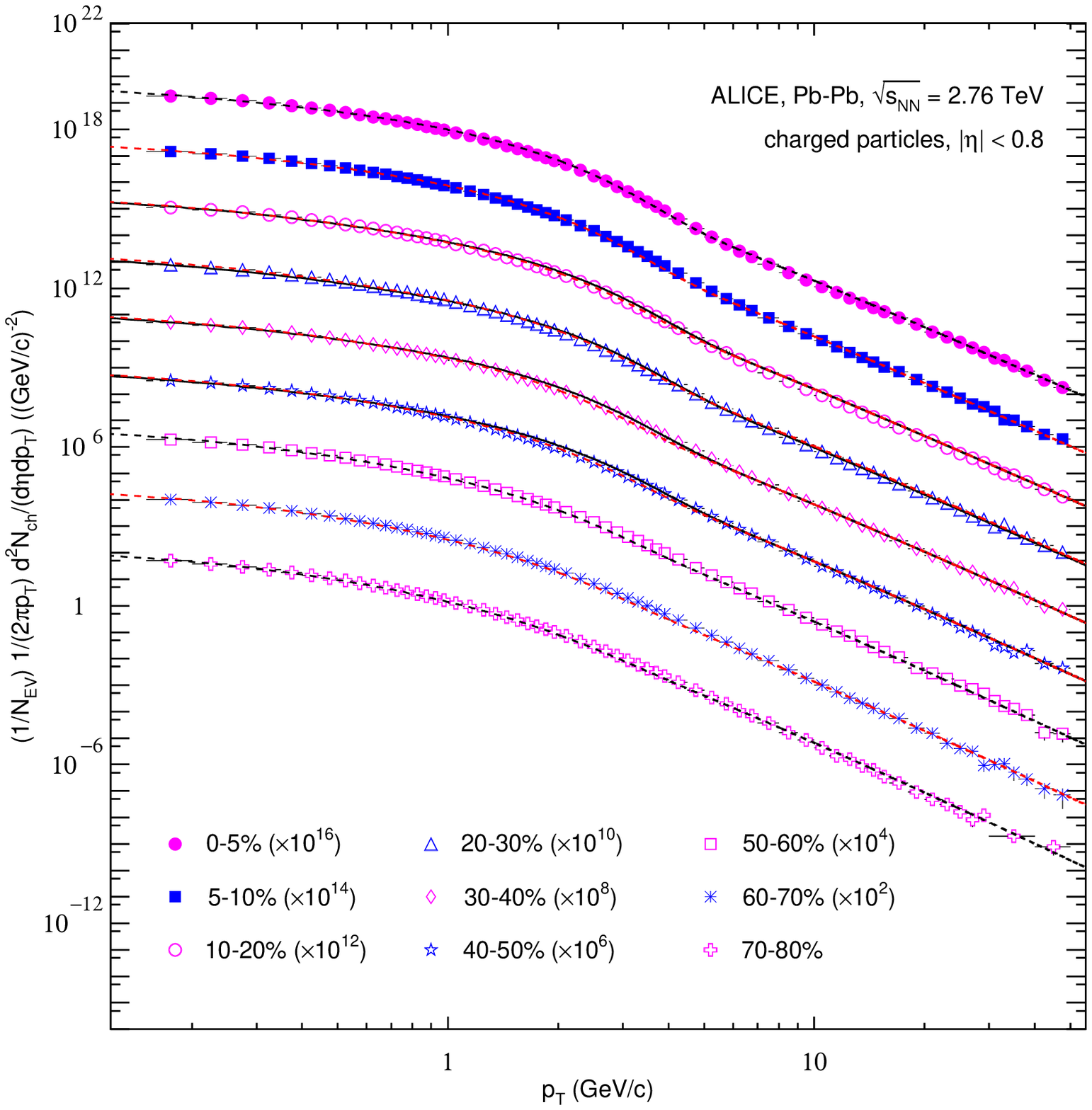}
\end{center}
\vskip.0cm Fig. 1. Double-differential spectra of charged
particles produced in Pb-Pb collisions at $\sqrt{s_{NN}}=2.76$ TeV
in the mid-pseudorapidity interval, $|\eta|<0.8$, for nine
centrality classes, 0--5\%, 5--10\%, 10--20\%, 20--30\%, 30--40\%,
40--50\%, 50--60\%, 60--70\%, and 70--80\%. The different
centrality classes are scaled down by different powers of ten for
plot clarity. The symbols represent the experimental data of the
ALICE Collaboration [25], and the solid curves for the
intermediate four cases (10--20\%, 20--30\%, 30--40\%, and
40--50\%) are our model results calculated by using the
superposition of the inverse power-law and revised Erlang
distribution due to Fig. 2. For comparison, the model results
correspond to the superposition of the inverse power-law and
(unrevised) Erlang distribution are displayed by the dashed
curves.
\end{figure*}

\begin{figure*}
\vskip.0cm \begin{center}
\includegraphics[width=15.cm]{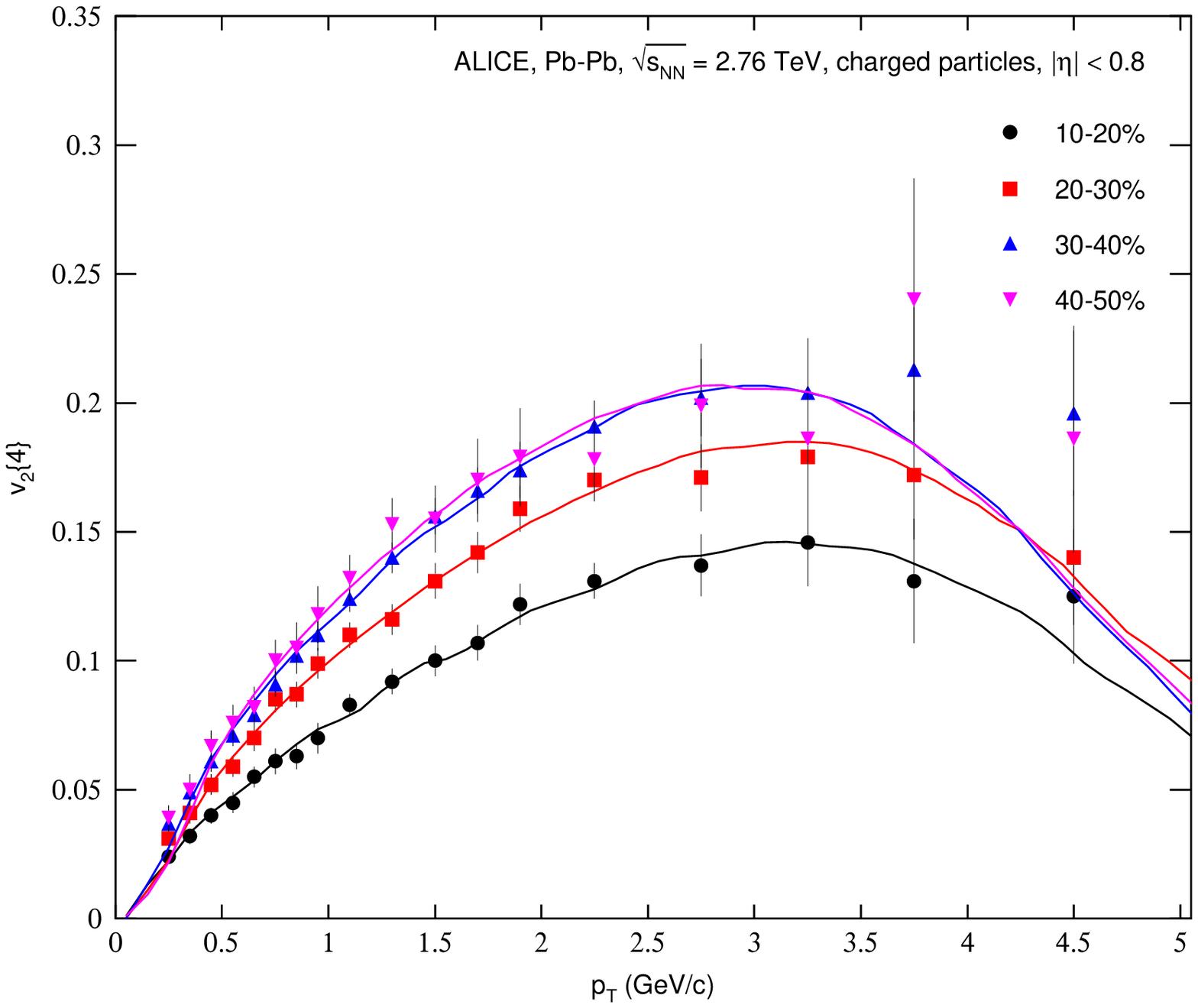}
\end{center}
\vskip.0cm Fig. 2. Dependence of elliptic flow on transverse
momentum for charged particles produced in Pb-Pb collisions at
$\sqrt{s_{NN}}=2.76$ TeV in $|\eta|<0.8$ for four centrality
classes, 10--20\%, 20--30\%, 30--40\%, and 40--50\%. The symbols
represent the experimental data of the ALICE Collaboration [26],
and the curves are our model results which are resulted from the
inverse power-paw which contributes zero elliptic flow and the
revised Erlang distribution which show anisotropic flow.
\end{figure*}

\begin{table*}
{\scriptsize Table 1. Values of free parameters ($p_{0}$, $n$,
$k_1$, $m$, and $\langle p_{Ti} \rangle$) and normalization
constant ($N_{p_T}$) for both the solid and dashed curves,
$\chi^2$ and dof$|_1$ for the solid curves, as well as $\chi^2$
and dof$|_2$ for the dashed curves in Figs. 1 and 3, where the
values of $m$ for Figs. 1, 3(a), 3(b), and 3(c) are invariably
taken to be 2, 2, 2, and 3 respectively, which are not listed in
the columns, and the values of $a_{x,y}$ and $b_{x,y}$ are the
same as Fig. 2.
\begin{center}
\begin{tabular}{ccccccccc}
\hline\hline Figure & Centrality & $p_{0}$ (GeV/$c$) & $n$ & $k_1$ & $\langle p_{Ti} \rangle$ (GeV/$c$) & $N_{p_T}$ & $\chi^2$/dof$|_1$ &$\chi^2$/dof$|_2$\\
\hline
Fig. 1    & 0--5\%   & $0.64\pm0.03$ & $6.10\pm0.31$ & $0.40\pm0.02$ & $0.38\pm0.01$ & $(2.30\pm0.12)\times10^{4}$ & $-$ & $112.114/58$ \\
            & 5--10\%  & $0.64\pm0.03$ & $6.12\pm0.31$ & $0.41\pm0.02$ & $0.37\pm0.01$ & $(2.00\pm0.10)\times10^{4}$ & $-$ & $101.732/58$\\
            & 10--20\% & $0.66\pm0.03$ & $6.15\pm0.31$ & $0.52\pm0.03$ & $0.39\pm0.01$ & $(1.40\pm0.07)\times10^{4}$ & $94.424/58$ & $149.176/58$\\
            & 20--30\% & $0.67\pm0.03$ & $6.19\pm0.31$ & $0.56\pm0.03$ & $0.39\pm0.01$ & $(9.70\pm0.48)\times10^{3}$ & $81.954/58$ & $136.474/58$\\
            & 30--40\% & $0.70\pm0.03$ & $6.24\pm0.31$ & $0.54\pm0.03$ & $0.37\pm0.01$ & $(6.50\pm0.33)\times10^{3}$ & $72.326/58$ & $164.256/58$\\
            & 40--50\% & $0.73\pm0.04$ & $6.31\pm0.32$ & $0.60\pm0.03$ & $0.38\pm0.01$ & $(3.80\pm0.19)\times10^{3}$ & $103.936/58$ & $162.214/58$\\
            & 50--60\% & $0.77\pm0.04$ & $6.50\pm0.33$ & $0.70\pm0.04$ & $0.37\pm0.01$ & $(2.40\pm0.12)\times10^{3}$ & $-$ & $109.214/58$\\
            & 60--70\% & $0.77\pm0.04$ & $6.50\pm0.33$ & $0.75\pm0.04$ & $0.36\pm0.01$ & $(1.20\pm0.06)\times10^{3}$ & $-$ & $136.358/58$\\
            & 70--80\% & $0.81\pm0.04$ & $6.60\pm0.33$ & $0.80\pm0.04$ & $0.36\pm0.01$ & $(5.50\pm0.28)\times10^{2}$ & $-$ & $82.012/58$\\
\hline
Fig. 3(a) & 0--5\%   & $0.56\pm0.03$ & $6.10\pm0.31$ & $0.48\pm0.02$ & $0.35\pm0.01$ & $(1.90\pm0.10)\times10^{4}$ & $349.944/56$ & $295.400/56$\\
            & 5--10\%  & $0.56\pm0.03$ & $6.12\pm0.31$ & $0.50\pm0.02$ & $0.35\pm0.01$ & $(1.50\pm0.08)\times10^{4}$ & $435.176/56$ & $455.056/56$\\
            & 10--20\% & $0.57\pm0.03$ & $6.15\pm0.31$ & $0.55\pm0.03$ & $0.35\pm0.01$ & $(1.20\pm0.06)\times10^{4}$ & $445.928/56$ & $383.432/56$\\
            & 20--30\% & $0.57\pm0.03$ & $6.19\pm0.31$ & $0.66\pm0.03$ & $0.35\pm0.01$ & $(8.70\pm0.44)\times10^{3}$ & $472.864/56$ & $646.072/56$\\
            & 30--40\% & $0.58\pm0.03$ & $6.22\pm0.31$ & $0.70\pm0.03$ & $0.36\pm0.01$ & $(5.40\pm0.27)\times10^{3}$ & $736.064/56$ & $971.992/56$\\
            & 40--50\% & $0.61\pm0.03$ & $6.30\pm0.32$ & $0.71\pm0.03$ & $0.35\pm0.01$ & $(3.40\pm0.17)\times10^{3}$ & $596.792/56$ & $789.488/56$\\
            & 40--60\% & $0.63\pm0.03$ & $6.40\pm0.32$ & $0.72\pm0.03$ & $0.35\pm0.01$ & $(2.90\pm0.14)\times10^{3}$ & $383.488/56$  & $249.312/56$\\
            & 60--80\% & $0.68\pm0.03$ & $6.50\pm0.33$ & $0.75\pm0.04$ & $0.35\pm0.01$ & $(8.00\pm0.40)\times10^{2}$ & $-$      & $103.208/56$\\
\hline
Fig. 3(b) & 0--5\%   & $1.14\pm0.06$ & $6.25\pm0.31$ & $0.10\pm0.01$ & $0.39\pm0.01$ & $3800\pm190$ & $128.367/51$ & $171.768/51$\\
            & 5--10\%  & $1.14\pm0.06$ & $6.25\pm0.31$ & $0.12\pm0.01$ & $0.39\pm0.01$ & $3100\pm155$ & $131.682/51$ & $202.521/51$\\
            & 10--20\% & $1.19\pm0.06$ & $6.35\pm0.32$ & $0.15\pm0.01$ & $0.38\pm0.01$ & $2200\pm110$ & $114.648/51$ & $382.449/51$\\
            & 20--30\% & $1.19\pm0.06$ & $6.35\pm0.32$ & $0.18\pm0.01$ & $0.38\pm0.01$ & $1400\pm85$  & $142.188/51$ & $489.192/51$\\
            & 30--40\% & $1.19\pm0.06$ & $6.40\pm0.32$ & $0.20\pm0.01$ & $0.37\pm0.01$ & $950\pm48$   & $458.286/51$ & $473.025/51$\\
            & 40--50\% & $1.19\pm0.06$ & $6.40\pm0.32$ & $0.22\pm0.01$ & $0.36\pm0.01$ & $590\pm30$   & $351.543/51$ & $494.853/51$\\
            & 40--60\% & $1.19\pm0.06$ & $6.40\pm0.32$ & $0.23\pm0.01$ & $0.35\pm0.01$ & $450\pm23$   & $177.786/51$ & $261.222/51$\\
            & 60--80\% & $1.81\pm0.09$ & $7.30\pm0.36$ & $0.22\pm0.01$ & $0.39\pm0.01$ & $110\pm6$    & $-$     & $84.099/51$\\
\hline
Fig. 3(c) & 0--5\%   & $1.50\pm0.08$ & $6.75\pm0.34$ & $0.07\pm0.01$ & $0.41\pm0.01$ & $1100\pm55$  & $136.458/42$ & $439.152/42$\\
            & 5--10\%  & $1.66\pm0.08$ & $7.05\pm0.35$ & $0.07\pm0.01$ & $0.40\pm0.01$ & $920\pm46$   & $125.244/42$ & $594.174/42$\\
            & 10--20\% & $1.70\pm0.09$ & $7.05\pm0.34$ & $0.07\pm0.01$ & $0.38\pm0.01$ & $720\pm36$   & $72.240/42$ & $1018.668/42$\\
            & 20--30\% & $1.70\pm0.09$ & $7.05\pm0.34$ & $0.08\pm0.01$ & $0.37\pm0.01$ & $500\pm25$   & $183.372/42$ & $1132.950/42$\\
            & 30--40\% & $1.89\pm0.10$ & $7.45\pm0.38$ & $0.11\pm0.01$ & $0.36\pm0.01$ & $310\pm16$   & $159.180/42$ & $1145.382/42$\\
            & 40--50\% & $2.02\pm0.10$ & $7.75\pm0.39$ & $0.14\pm0.01$ & $0.36\pm0.01$ & $180\pm9$    & $235.536/42$ & $929.082/42$\\
            & 40--60\% & $2.02\pm0.10$ & $7.75\pm0.39$ & $0.15\pm0.01$ & $0.36\pm0.01$ & $150\pm7$    & $175.686/42$ & $327.474/42$\\
            & 60--80\% & $2.09\pm0.10$ & $7.75\pm0.39$ & $0.15\pm0.01$ & $0.36\pm0.01$ & $45\pm2$     & $-$ & $44.730/42$\\
\hline
\end{tabular}%
\end{center}}
\end{table*}

\newpage
\begin{table*}
{\scriptsize Table 2. Values of free parameters ($a_{x}$, $b_{x}$,
$a_{y}$, and $b_{y}$), $\chi^2$, and dof corresponding to the
curves in Figs. 2 and 4. These parameters consists of the revised
Erlang distribution. The parameters $b_x$ and $b_y$ can take
simultaneously positive and negative values. Both the
probabilities for positive and negative $b_{x,y}$ are 50\%. The
solid curves contained the revised Erlang distribution in Fig. 1
also use the same parameters as Fig. 2.
\begin{center}
\begin{tabular}{ccccccc}
\hline\hline Figure & Centrality & $a_{x}$ & $\pm b_{x}$ & $a_{y}$ & $\pm b_{y}$ & $\chi^2$/dof \\
\hline
Fig. 2    & 10--20\% & $1.09\pm0.01$ & $0.23\pm0.01$ & $1.00\pm0.01$ & $0.00\pm0.01$ & $5.304/8$ \\
            & 20--30\% & $1.11\pm0.01$ & $0.29\pm0.01$ & $1.00\pm0.01$ & $0.00\pm0.01$ & $13.632/8$ \\
            & 30--40\% & $1.14\pm0.01$ & $0.30\pm0.01$ & $1.00\pm0.01$ & $0.00\pm0.01$ & $20.488/8$ \\
            & 40--50\% & $1.14\pm0.01$ & $0.35\pm0.01$ & $1.00\pm0.01$ & $0.00\pm0.01$ & $21.136/8$ \\
\hline
Fig. 4(a) & 0--5\%   & $1.04\pm0.01$ & $0.10\pm0.01$ & $1.00\pm0.01$ & $0.00\pm0.01$ & $150.162/29$ \\
            & 5--10\%  & $1.06\pm0.01$ & $0.20\pm0.01$ & $1.00\pm0.01$ & $0.00\pm0.01$ & $162.980/29$ \\
            & 10--20\% & $1.10\pm0.01$ & $0.24\pm0.01$ & $1.00\pm0.01$ & $0.00\pm0.01$ & $248.675/29$ \\
            & 20--30\% & $1.12\pm0.01$ & $0.35\pm0.01$ & $1.00\pm0.01$ & $0.00\pm0.01$ & $287.419/29$ \\
            & 30--40\% & $1.12\pm0.01$ & $0.42\pm0.01$ & $1.00\pm0.01$ & $0.00\pm0.01$ & $449.500/29$ \\
            & 40--50\% & $1.13\pm0.01$ & $0.43\pm0.01$ & $1.00\pm0.01$ & $0.00\pm0.01$ & $472.178/29$ \\
            & 50--60\% & $1.13\pm0.01$ & $0.43\pm0.01$ & $1.00\pm0.01$ & $0.00\pm0.01$ & $462.115/29$ \\
\hline
Fig. 4(b) & 0--5\%   & $1.05\pm0.01$ & $0.03\pm0.01$ & $1.00\pm0.01$ & $0.01\pm0.01$ & $87.494/22$ \\
            & 5--10\%  & $1.09\pm0.01$ & $0.04\pm0.01$ & $1.00\pm0.01$ & $0.01\pm0.01$ & $130.438/22$ \\
            & 10--20\% & $1.14\pm0.01$ & $0.06\pm0.01$ & $1.00\pm0.01$ & $0.01\pm0.01$ & $184.734/22$ \\
            & 20--30\% & $1.19\pm0.01$ & $0.11\pm0.01$ & $1.00\pm0.01$ & $0.09\pm0.01$ & $232.232/22$ \\
            & 30--40\% & $1.18\pm0.01$ & $0.45\pm0.01$ & $1.00\pm0.01$ & $0.33\pm0.01$ & $197.318/22$ \\
            & 40--50\% & $1.18\pm0.01$ & $0.44\pm0.01$ & $1.00\pm0.01$ & $0.31\pm0.01$ & $177.870/22$ \\
            & 50--60\% & $1.16\pm0.01$ & $0.47\pm0.01$ & $1.00\pm0.01$ & $0.33\pm0.01$ & $130.460/22$ \\
\hline
Fig. 4(c) & 0--5\%   & $1.05\pm0.01$ & $0.33\pm0.01$ & $1.00\pm0.01$ & $0.30\pm0.01$ & $73.467/27$ \\
            & 5--10\%  & $1.08\pm0.01$ & $0.36\pm0.01$ & $1.00\pm0.01$ & $0.30\pm0.01$ & $101.142/27$ \\
            & 10--20\% & $1.10\pm0.01$ & $0.49\pm0.01$ & $1.00\pm0.01$ & $0.36\pm0.01$ & $131.679/27$ \\
            & 20--30\% & $1.11\pm0.01$ & $0.57\pm0.02$ & $1.00\pm0.01$ & $0.36\pm0.01$ & $151.065/27$ \\
            & 30--40\% & $1.13\pm0.01$ & $0.59\pm0.02$ & $1.00\pm0.01$ & $0.35\pm0.01$ & $159.813/27$ \\
            & 40--50\% & $1.13\pm0.01$ & $0.60\pm0.02$ & $1.00\pm0.01$ & $0.30\pm0.01$ & $144.234/27$ \\
            & 50--60\% & $1.11\pm0.01$ & $0.60\pm0.02$ & $1.00\pm0.01$ & $0.26\pm0.01$ & $186.219/27$ \\
\hline
\end{tabular}%
\end{center}}
\end{table*}

\begin{figure*}
\vskip.0cm \begin{center}
\includegraphics[width=15.cm]{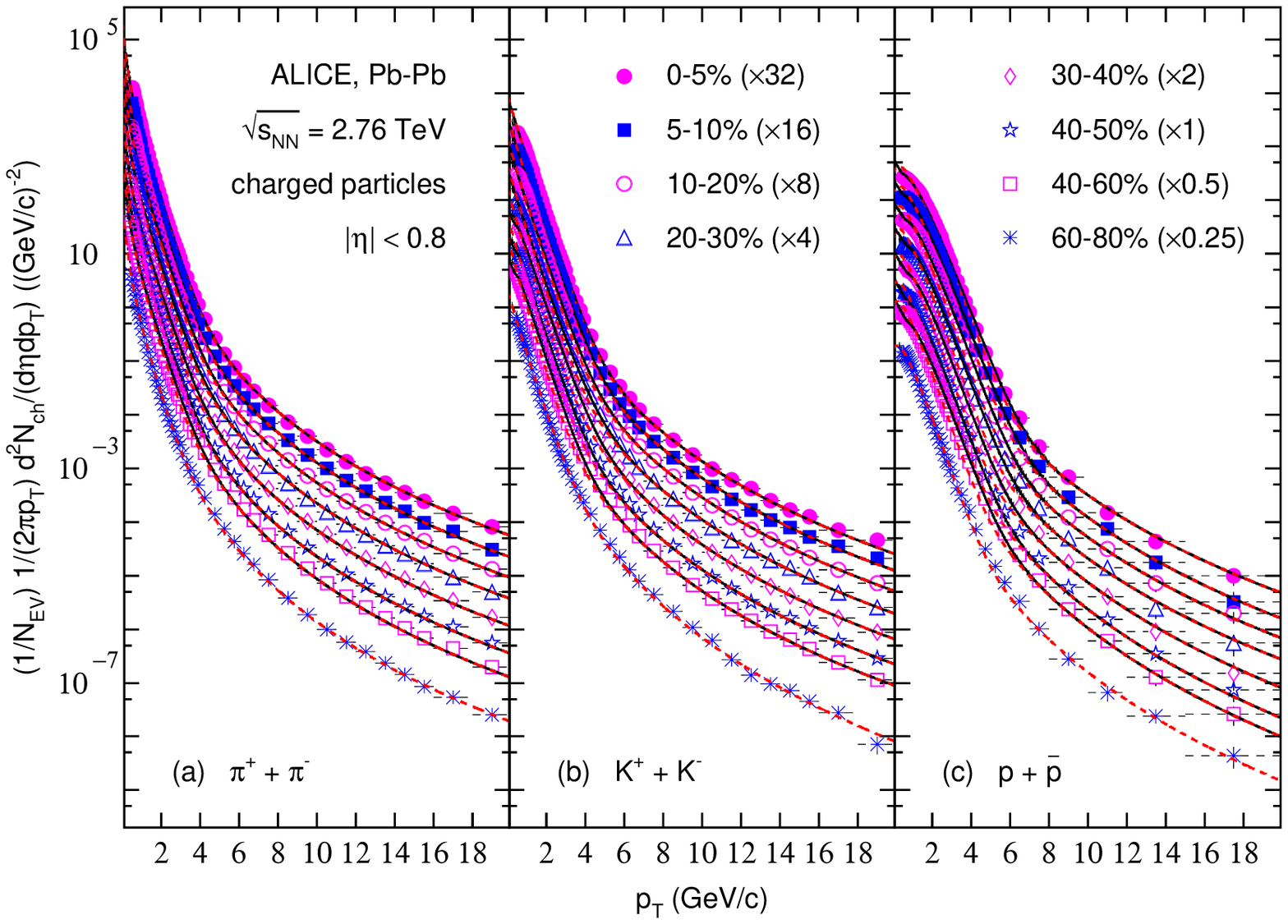}
\end{center}
\vskip.0cm Fig. 3. Double-differential spectra of charged (a)
pions ($\pi^++\pi^-$), (b) kaons ($K^++K^-$), and (c)
(anti)protons ($p+\bar p$) produced in Pb-Pb collisions at
$\sqrt{s_{NN}}=2.76$ TeV in $|\eta|<0.8$ for eight centrality
classes, 0--5\%, 5--10\%, 10--20\%, 20--30\%, 30--40\%, 40--50\%,
40--60\%, and 60--80\%. The different centrality classes are
scaled down by different factors listed in the panel for plot
clarity. The symbols represent the experimental data of the ALICE
Collaboration [27] and the solid curves for the first seven cases
are our model results due to Fig. 4. For comparison, the model
results correspond to the superposition of the inverse power-law
and (unrevised) Erlang distribution are displayed by the dashed
curves.
\end{figure*}

%\newpage
\begin{figure*}
\vskip.0cm \begin{center}
\includegraphics[width=15.cm]{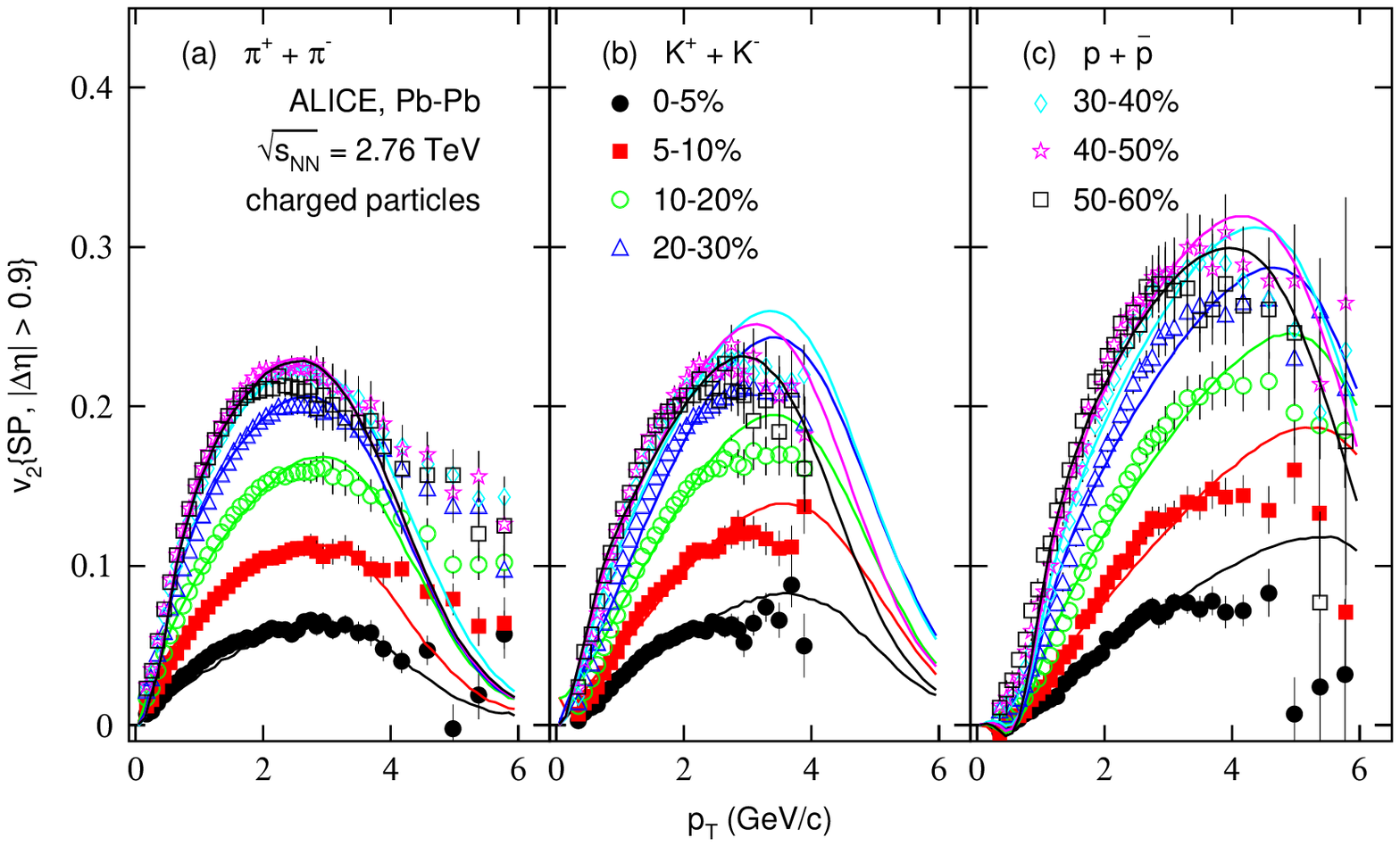}
\end{center}
\vskip.0cm Fig. 4. Dependence of elliptic flow on transverse
momentum for (a) $\pi^++\pi^-$, (b) $K^++K^-$, and (c) $p+\bar p$
produced in Pb-Pb collisions at $\sqrt{s_{NN}}=2.76$ TeV in
$|\eta|<0.8$ for seven centrality classes, 0--5\%, 5--10\%,
10--20\%, 20--30\%, 30--40\%, 40--50\%, and 50--60\%. The symbols
represent the experimental data of the ALICE Collaboration [28]
and the curves are our model results contained the revised Erlang
distribution.
\end{figure*}

Figure 5 gives the $\eta$ spectra of charged particles produced in
Pb-Pb collisions at $\sqrt{s_{NN}}=2.76$ TeV in ten centrality
classes, 0--5\%, 5--10\%, 10--20\%, 20--30\%, 30--40\%, 40--50\%,
50--60\%, 60--70\%, 70--80\%, and 80--90\%. The symbols represent
the experimental data of the ALICE Collaborations [29--31], where
the pink, black, blue, and green symbols represent the
measurements from the Silicon Pixel Detector in refs. [29] and
[30], as well as from the Forward Multiplicity Detector in refs.
[29] and [31], respectively. The solid curves are our results
fitted by the three-Gaussian function, and the dashed curves are
our results fitted by another set of parameters in the
three-Gaussian function for the purpose of comparison with the
solid curves. The first, second, and third Gaussian functions
describe the contributions of backward, central, and forward
sources (or regions or components) in the $\eta$ space
respectively. The values of free parameter ($\delta \eta$,
$\sigma_{\eta1}$, $\sigma_{\eta2}$, and $k_2$), normalization
constant ($N_\eta$), $\chi^2$, and dof fitted by us are listed in
the upper and lower panels in Table 3 for the solid and dashed
curves respectively. One can see that the model results with the
two sets of parameters describe approximately the ALICE
experimental $\eta$ spectra of charged particles measured in
different centrality classes in Pb-Pb collisions at
$\sqrt{s_{NN}}=2.76$ TeV. The $\eta$ spectra obtained in different
centrality classes at TeV energy confirm the three-source or other
non-sole source model in high energy collisions [41--52], though
the tendencies of parameters $\sigma_{\eta1}$ and $k_2$ are
optional, where $\sigma_{\eta1}$ is invariant or increased and
$k_2$ is decreased or invariant with increasing the centrality
percentage. In the case of having more data in the backward
(forward) $\eta$ region, we can obtain the parameters as
accurately as possible.

%\newpage
\begin{figure*}
\vskip.0cm \begin{center}
\includegraphics[width=15.cm]{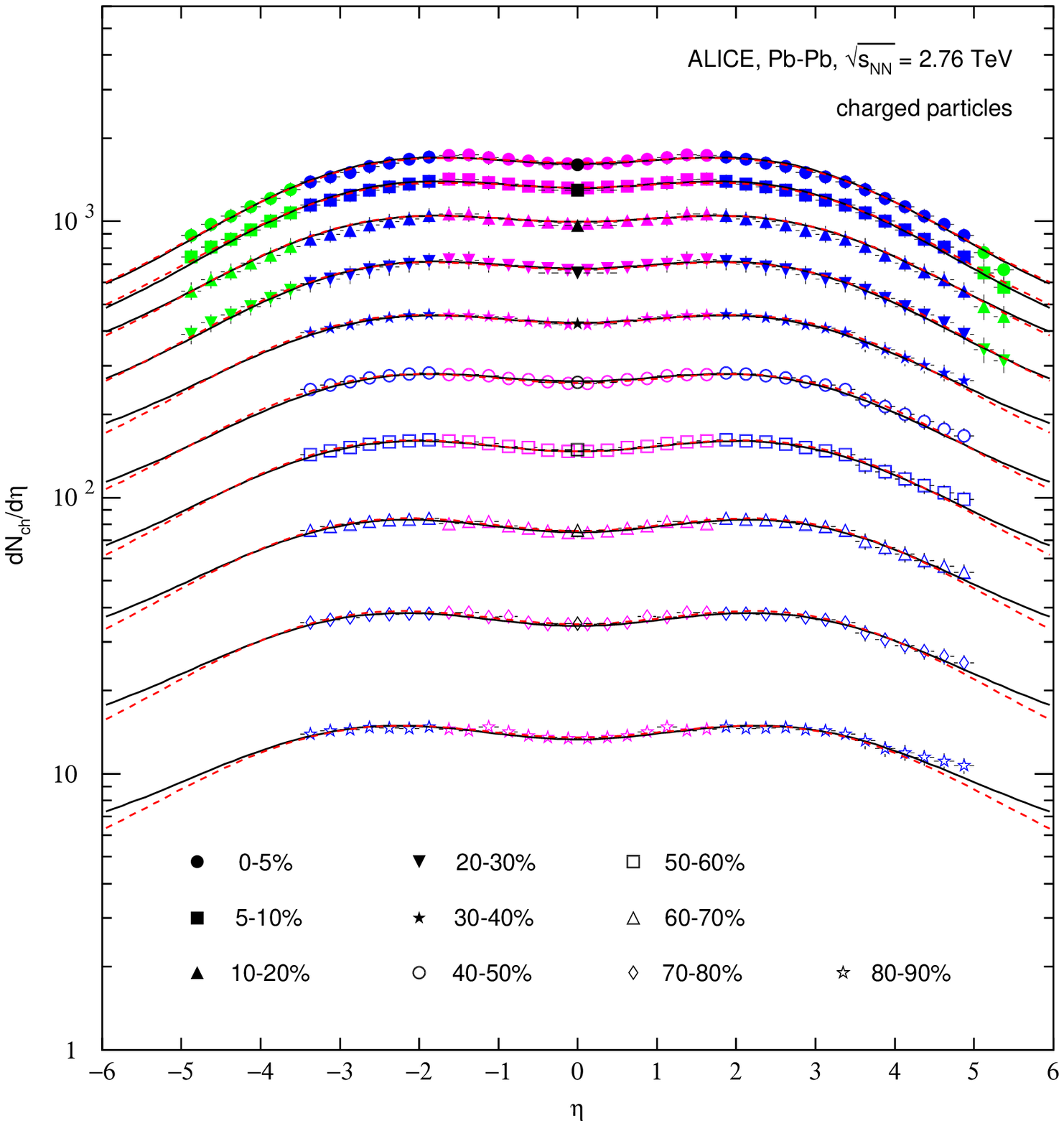}
\end{center}
\vskip.0cm  Fig. 5. Pseudorapidity spectra of charged particles
produced in Pb-Pb collisions at $\sqrt{s_{NN}}=2.76$ TeV in ten
centrality classes, 0--5\%, 5--10\%, 10--20\%, 20--30\%, 30--40\%,
40--50\%, 50--60\%, 60--70\%, 70--80\%, and 80--90\%. The symbols
represent the experimental data of the ALICE Collaborations
[29--31], where the pink, black, blue, and green symbols represent
the measurements from the Silicon Pixel Detector in refs. [29] and
[30], as well as from the Forward Multiplicity Detector in refs.
[29] and [31], respectively. The solid and dashed curves are our
results fitted by the three-Gaussian function with two different
sets of parameters.
\end{figure*}

Based on the parameter values obtained from Figs. 1, 2, and 5 and
listed in Tables 1--3, we can perform the Monte Carlo calculation
and obtain the values of a series of kinematical quantities. Based
on these kinematical quantities, we can structure some scatter
plots of charged particles at the kinetic freeze-out, and these
scatter plots reflect the event patterns at the last stage of the
interaction process. As an example, the following discussions are
only based on the upper panel of Table 3 in the case of Table 3
being used. The result based on the lower panel of Table 3 is not
presented due to its triviality. In fact, what we parameterize for
the $p_T$ and $\eta$ (or $Y$) spectra in the above is independent
of models. We can even use the experimental discrete values
themselves to replace these parameterizations. However, the
experimental data are often measured in partly phase space. This
is not enough for us to structure fully the event patterns. It is
necessary to use a model for extrapolation elsewhere.

Figure 6 gives the event patterns displayed by the scatter plots
of charged particles in the three-dimensional
$\beta_x-\beta_y-\beta_z$ space. The panels (a)(b), (c)(d),
(e)(f), and (g)(h) correspond to the results for the centrality
classes 10--20\%, 20--30\%, 30--40\%, and 40--50\%, respectively.
At the same time, the left (right) panel corresponds to the
results contained the revised (unrevised) Erlang distribution. The
blue and red globules represent the contributions of inverse
power-law and revised (unrevised) Erlang distribution
respectively. The total number of charged particles for each panel
is 1000. The values of root-mean-squares
$\sqrt{\overline{\beta_x^2}}$ for $\beta_x$,
$\sqrt{\overline{\beta_y^2}}$ for $\beta_y$, and
$\sqrt{\overline{\beta_z^2}}$ for $\beta_z$, as well as the
maximum $|\beta_x|$, $|\beta_y|$, and $|\beta_z|$ (i.e.
$|\beta_x|_{\max}$, $|\beta_y|_{\max}$, and $|\beta_z|_{\max}$)
are listed in Table 4. One can see that, for the four centrality
classes, the event patterns displayed by the scatter plots of
charged particles in the three-dimensional
$\beta_x-\beta_y-\beta_z$ space are rough sphericity (or fat
ellipsoid along the $Oz$ axis) with high density close to
$\beta_z=1$. In particular, $\sqrt{\overline{\beta_y^2}} \leq
\sqrt{\overline{\beta_x^2}} < \sqrt{\overline{\beta_z^2}}$ and
$|\beta_y|_{\max} \approx |\beta_x|_{\max} \approx
|\beta_z|_{\max}$.

Figure 7 is the same as Fig. 6, but it shows the event patterns in
the three-dimensional $p_x-p_y-p_z$ space. The values of
root-mean-squares $\sqrt{\overline{p_x^2}}$ for $p_x$,
$\sqrt{\overline{p_y^2}}$ for $p_y$, and $\sqrt{\overline{p_z^2}}$
for $p_z$, as well as the maximum $|p_x|$, $|p_y|$, and $|p_z|$
(i.e. $|p_x|_{\max}$, $|p_y|_{\max}$, and $|p_z|_{\max}$) are
listed in Table 5. One can see that, for the four centrality
classes, if the relative large size in the $p_z$ direction is
considered, the event patterns in the three-dimensional
$p_x-p_y-p_z$ space can be regarded as rough cylinders with some
removed particles from the profile. In particular,
$\sqrt{\overline{p_y^2}} \leq \sqrt{\overline{p_x^2}} \ll
\sqrt{\overline{p_z^2}}$ and $|p_y|_{\max} \approx |p_x|_{\max}
\ll |p_z|_{\max}$.

Figure 8 is also the same as Fig. 6, but it shows the event
patterns in the three-dimensional $Y_1-Y_2-Y$ space. The values of
root-mean-squares $\sqrt{\overline{Y_1^2}}$ for $Y_1$,
$\sqrt{\overline{Y_2^2}}$ for $Y_2$, and $\sqrt{\overline{Y^2}}$
for $Y$, as well as the maximum $|Y_1|$, $|Y_2|$, and $|Y|$ (i.e.
$|Y_1|_{\max}$, $|Y_2|_{\max}$, and $|Y|_{\max}$) are listed in
Table 6. One can see that, for the four centrality classes, the
event patterns in the three-dimensional $Y_1-Y_2-Y$ space are
rough cylinder with a high peak at the top and a long tail at the
bottom. In particular, $\sqrt{\overline{Y_2^2}} \leq
\sqrt{\overline{Y_1^2}} < \sqrt{\overline{Y^2}}$ and $|Y_2|_{\max}
\approx |Y_1|_{\max} < |Y|_{\max}$.

\begin{figure*}
\vskip-1.0cm \begin{center}
\includegraphics[width=13.cm]{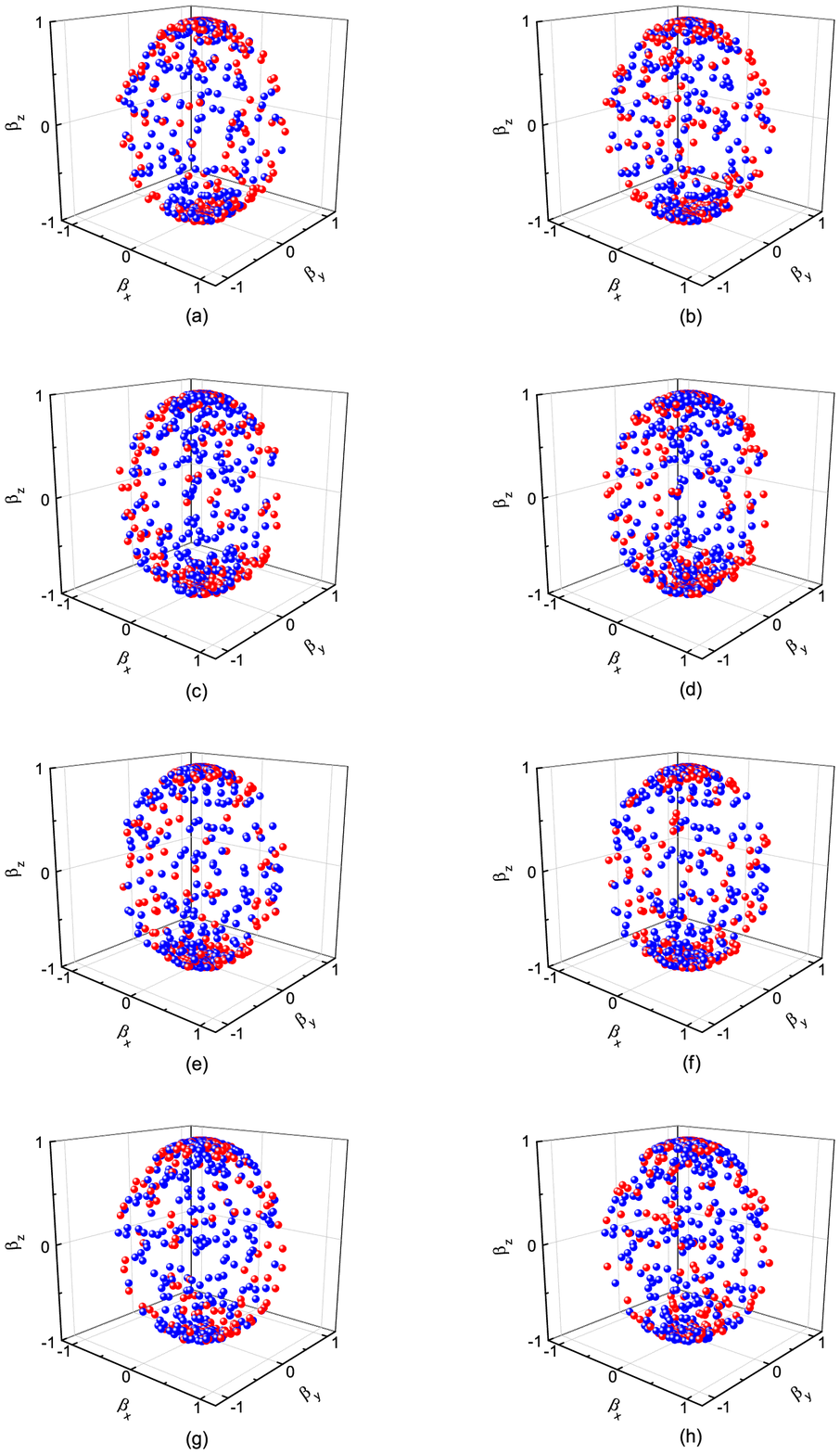}
\end{center}
\vskip.0cm Fig. 6. Event patterns displayed by the scatter plots
of charged particles in three-dimensional velocity
($\beta_x-\beta_y-\beta_z$) space in Pb-Pb collisions at
$\sqrt{s_{NN}}=2.76$ TeV in four centrality classes (a)(b)
10--20\%, (c)(d) 20--30\%, (e)(f) 30--40\%, and (g)(h) 40--50\%.
The number of charged particles for each panel is 1000. The blue
and red globules in the left panel represent the contributions of
the inverse power-law and revised Erlang distribution for $p_T$
respectively, and those in the right panel correspond to the
contributions of the inverse power-law and (unrevised) Erlang
distribution. The blue globules presented in the left and right
panels are totally the same, and the red globules presented in
both the panels are similar to each other.
\end{figure*}

\begin{figure*}
\vskip.0cm \begin{center}
\includegraphics[width=13.cm]{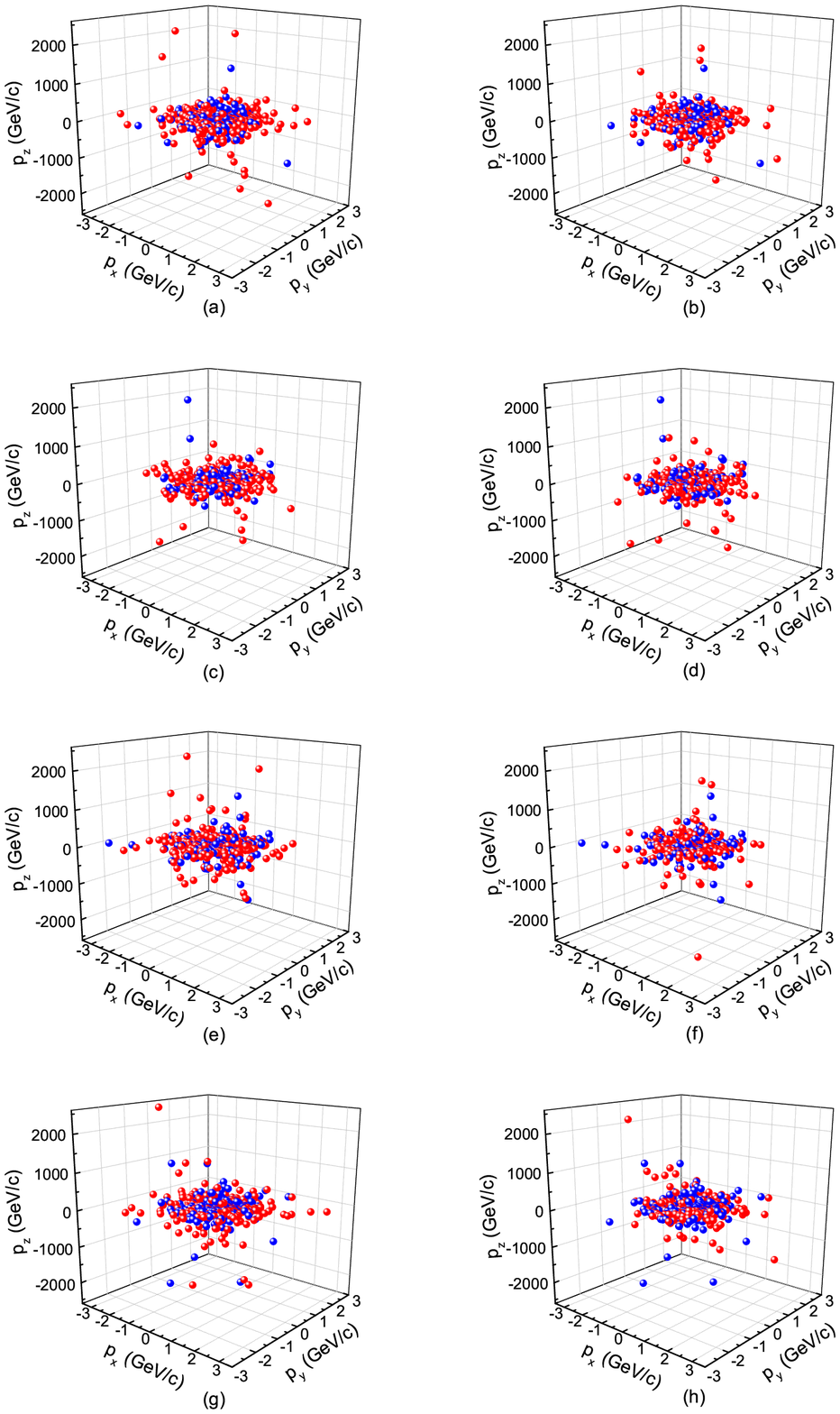}
\end{center}
\vskip.0cm Fig. 7. Same as Fig. 6, but showing the results in
three-dimensional momentum ($p_x-p_y-p_z$) space.
\end{figure*}

\begin{figure*}
\vskip.0cm \begin{center}
\includegraphics[width=13.cm]{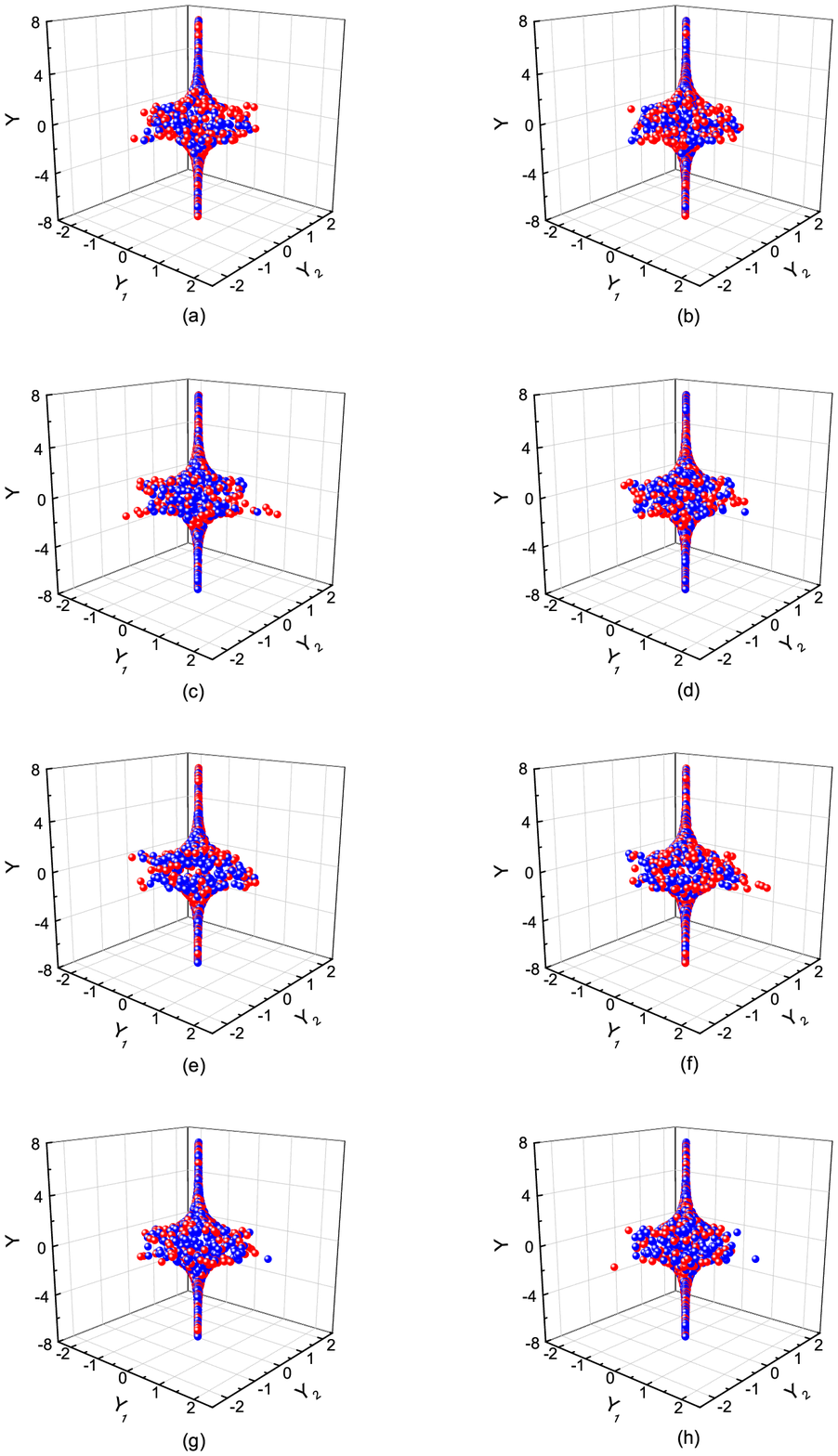}
\end{center}
\vskip.0cm Fig. 8. Same as Fig. 6, but showing the results in
three-dimensional rapidity ($Y_1-Y_2-Y$) space.
\end{figure*}

\vskip1.0cm
\begin{table*}
{\scriptsize Table 3. Upper panel: Values of free parameter
($\delta \eta$, $\sigma_{\eta1}$, $\sigma_{\eta2}$, and $k_2$),
normalization constant ($N_\eta$), $\chi^2$, and dof corresponding
to the solid curves in Fig. 5, where $\eta_C=0$ is not listed in
the column. Lower panel: Same as the upper panel, but showing the
values corresponding to the dashed curves in Fig. 5.
\begin{center}
\begin{tabular}{cccccccc}
\hline\hline Type & Centrality  & $\delta \eta$ & $\sigma_{\eta1}$ & $\sigma_{\eta2}$ & $k_2$ & $N_{\eta}$ & $\chi^2$/dof \\
\hline
Solid  & 0--5\%   & $2.18\pm0.11$ & $1.80\pm0.09$  & $8.00\pm1.00$  & $0.21\pm0.01$ & $(1.60\pm0.08)\times10^{5}$ & $6.462/37$ \\
Curves & 5--10\%  & $2.18\pm0.11$ & $1.80\pm0.09$  & $8.00\pm1.00$  & $0.21\pm0.01$ & $(1.31\pm0.07)\times10^{5}$ & $5.360/37$ \\
       & 10--20\% & $2.20\pm0.11$ & $1.82\pm0.09$  & $8.60\pm1.00$  & $0.19\pm0.01$ & $(1.00\pm0.05)\times10^{5}$ & $5.278/37$ \\
       & 20--30\% & $2.20\pm0.11$ & $1.80\pm0.09$  & $8.60\pm1.00$  & $0.19\pm0.01$ & $(6.80\pm0.34)\times10^{4}$ & $4.007/37$ \\
       & 30--40\% & $2.22\pm0.11$ & $1.80\pm0.09$  & $9.00\pm1.00$  & $0.17\pm0.01$ & $(4.40\pm0.22)\times10^{4}$ & $3.535/29$ \\
       & 40--50\% & $2.22\pm0.11$ & $1.80\pm0.09$  & $9.10\pm1.00$  & $0.17\pm0.01$ & $(2.70\pm0.14)\times10^{4}$ & $5.485/29$ \\
       & 50--60\% & $2.32\pm0.12$ & $1.82\pm0.09$  & $10.00\pm1.00$ & $0.17\pm0.01$ & $(1.56\pm0.08)\times10^{4}$ & $3.786/29$ \\
       & 60--70\% & $2.40\pm0.12$ & $1.82\pm0.09$  & $10.00\pm1.00$ & $0.16\pm0.01$ & $(8.20\pm0.41)\times10^{3}$ & $6.990/29$ \\
       & 70--80\% & $2.43\pm0.12$ & $1.82\pm0.09$  & $10.00\pm1.00$ & $0.15\pm0.01$ & $(3.80\pm0.19)\times10^{3}$ & $9.934/29$ \\
       & 80--90\% & $2.48\pm0.13$ & $1.82\pm0.09$  & $10.00\pm1.00$ & $0.14\pm0.01$ & $(1.50\pm0.07)\times10^{3}$ & $9.909/28$ \\
\hline
Dashed & 0--5\%   & $2.17\pm0.11$ & $1.84\pm0.09$  & $8.00\pm1.00$  & $0.21\pm0.01$ & $(1.60\pm0.08)\times10^{5}$ & $5.388/37$ \\
Curves & 5--10\%  & $2.20\pm0.11$ & $1.86\pm0.09$  & $8.00\pm1.00$  & $0.21\pm0.01$ & $(1.31\pm0.07)\times10^{5}$ & $3.763/37$ \\
       & 10--20\% & $2.24\pm0.11$ & $1.86\pm0.09$  & $8.00\pm1.00$  & $0.21\pm0.01$ & $(1.00\pm0.05)\times10^{5}$ & $6.092/37$ \\
       & 20--30\% & $2.27\pm0.11$ & $1.86\pm0.09$  & $8.00\pm1.00$  & $0.21\pm0.01$ & $(6.80\pm0.34)\times10^{4}$ & $3.306/37$ \\
       & 30--40\% & $2.30\pm0.11$ & $1.86\pm0.09$  & $9.00\pm1.00$  & $0.21\pm0.01$ & $(4.40\pm0.22)\times10^{4}$ & $4.020/29$ \\
       & 40--50\% & $2.33\pm0.11$ & $1.88\pm0.09$  & $9.00\pm1.00$  & $0.21\pm0.01$ & $(2.70\pm0.14)\times10^{4}$ & $3.759/29$ \\
       & 50--60\% & $2.38\pm0.12$ & $1.88\pm0.09$  & $10.00\pm1.00$ & $0.21\pm0.01$ & $(1.56\pm0.08)\times10^{4}$ & $3.880/29$ \\
       & 60--70\% & $2.44\pm0.12$ & $1.90\pm0.09$  & $10.00\pm1.00$ & $0.21\pm0.01$ & $(8.20\pm0.41)\times10^{3}$ & $7.169/29$ \\
       & 70--80\% & $2.47\pm0.12$ & $1.91\pm0.09$  & $10.00\pm1.00$ & $0.21\pm0.01$ & $(3.80\pm0.19)\times10^{3}$ & $7.760/29$ \\
       & 80--90\% & $2.51\pm0.13$ & $1.96\pm0.09$  & $10.00\pm1.00$ & $0.21\pm0.01$ & $(1.50\pm0.07)\times10^{3}$ & $15.671/28$ \\
\hline
\end{tabular}%
\end{center}}
\end{table*}

\vskip1.0cm
\begin{table*}
{\scriptsize Table 4. Values of the root-mean-squares
$\sqrt{\overline{\beta_x^2}}$ for $\beta_x$,
$\sqrt{\overline{\beta_y^2}}$ for $\beta_y$, and
$\sqrt{\overline{\beta_z^2}}$ for $\beta_z$, as well as the
maximum $|\beta_x|$, $|\beta_y|$, and $|\beta_z|$ (i.e.
$|\beta_x|_{\max}$, $|\beta_y|_{\max}$, and $|\beta_z|_{\max}$)
corresponding to the scatter plots in Figs. 6(a)(b)--6(g)(h) which
show 2.76 TeV Pb-Pb collisions with centrality classes 10--20\%,
20--30\%, 30--40\%, and 40--50\%, respectively. The upper panel in
the table is the results presented in the left panel in the figure
and contained the revised Erlang distribution, and the lower panel
in the table is the results presented in the right panel in the
figure and contained the (unrevised) Erlang distribution. Both the
root-mean-squares and the maximum velocity components are in the
units of $c$.
\begin{center}
\begin{tabular}{cccccccc}
\hline\hline Type & Centrality & $\sqrt{\overline{\beta_x^2}}$ & $\sqrt{\overline{\beta_y^2}}$ & $\sqrt{\overline{\beta_z^2}}$ & $|\beta_x|_{\max}$ & $|\beta_y|_{\max}$ & $|\beta_z|_{\max}$ \\
\hline
Revised   & 10--20\% & $0.264\pm0.009$ & $0.264\pm0.010$  & $0.897\pm0.005$  & $0.931$ & $0.976$ & $1.000$ \\
Erlang    & 20--30\% & $0.278\pm0.010$ & $0.263\pm0.009$  & $0.888\pm0.005$  & $0.983$ & $0.979$ & $1.000$ \\
          & 30--40\% & $0.278\pm0.010$ & $0.240\pm0.009$  & $0.901\pm0.005$  & $0.970$ & $0.949$ & $1.000$ \\
          & 40--50\% & $0.279\pm0.009$ & $0.264\pm0.009$  & $0.890\pm0.006$  & $0.963$ & $0.947$ & $1.000$ \\
\hline
Unrevised & 10--20\% & $0.254\pm0.009$ & $0.268\pm0.010$  & $0.895\pm0.005$  & $0.897$ & $0.987$ & $1.000$ \\
Erlang    & 20--30\% & $0.267\pm0.009$ & $0.268\pm0.009$  & $0.886\pm0.006$  & $0.964$ & $0.994$ & $1.000$ \\
          & 30--40\% & $0.269\pm0.010$ & $0.244\pm0.009$  & $0.901\pm0.005$  & $0.972$ & $0.954$ & $1.000$ \\
          & 40--50\% & $0.265\pm0.009$ & $0.271\pm0.009$  & $0.888\pm0.006$  & $0.963$ & $0.984$ & $1.000$ \\
\hline
\end{tabular}%
\end{center}}
\end{table*}

\newpage
\begin{table*}
{\scriptsize Table 5. Same as Table 4, but showing the values of
the root-mean-squares $\sqrt{\overline{p_x^2}}$ for $p_x$,
$\sqrt{\overline{p_y^2}}$ for $p_y$, and $\sqrt{\overline{p_z^2}}$
for $p_z$, as well as the maximum $|p_x|$, $|p_y|$, and $|p_z|$
(i.e. $|p_x|_{\max}$, $|p_y|_{\max}$, and $|p_z|_{\max}$)
corresponding to the scatter plots in Fig. 7. All the
root-mean-squares and the maximum momentum components are in the
units of GeV/$c$.
\begin{center}
\begin{tabular}{cccccccc}
\hline\hline Type & Centrality & $\sqrt{\overline{p_x^2}}$  & $\sqrt{\overline{p_y^2}}$  & $\sqrt{\overline{p_z^2}}$  & $|p_x|_{\max}$  & $|p_y|_{\max}$  & $|p_z|_{\max}$  \\
\hline
Revised   & 10--20\% & $0.626\pm0.028$ & $0.557\pm0.022$  & $210.9\pm25.2$   & $4.876$ & $2.807$ & $2444.7$ \\
Erlang    & 20--30\% & $0.613\pm0.024$ & $0.537\pm0.032$  & $163.2\pm19.6$   & $3.390$ & $5.340$ & $1967.4$ \\
          & 30--40\% & $0.620\pm0.022$ & $0.505\pm0.019$  & $197.6\pm21.1$   & $3.130$ & $2.292$ & $2237.6$ \\
          & 40--50\% & $0.637\pm0.024$ & $0.589\pm0.025$  & $222.2\pm23.9$   & $3.061$ & $3.554$ & $2583.9$ \\
\hline
Unrevised & 10--20\% & $0.567\pm0.025$ & $0.557\pm0.022$  & $197.3\pm22.5$   & $4.263$ & $2.807$ & $2219.9$ \\
Erlang    & 20--30\% & $0.540\pm0.022$ & $0.537\pm0.032$  & $152.1\pm20.2$   & $3.390$ & $5.340$ & $1967.4$ \\
          & 30--40\% & $0.546\pm0.021$ & $0.505\pm0.019$  & $168.4\pm17.2$   & $3.130$ & $2.292$ & $1736.7$ \\
          & 40--50\% & $0.547\pm0.020$ & $0.589\pm0.025$  & $204.2\pm20.5$   & $2.460$ & $3.554$ & $1965.5$ \\
\hline
\end{tabular}%
\end{center}}
\end{table*}

\vskip1.0cm
\begin{table*}
{\scriptsize Table 6. Same as Table 4, but showing the values of
the root-mean-squares $\sqrt{\overline{Y_1^2}}$ for $Y_1$,
$\sqrt{\overline{Y_2^2}}$ for $Y_2$, and $\sqrt{\overline{Y^2}}$
for $Y$, as well as the maximum $|Y_1|$, $|Y_2|$, and $|Y|$ (i.e.
$|Y_1|_{\max}$, $|Y_2|_{\max}$, and $|Y|_{\max}$) corresponding to
the scatter plots in Fig. 8.
\begin{center}
\begin{tabular}{cccccccc}
\hline\hline Type & Centrality & $\sqrt{\overline{Y_1^2}}$  & $\sqrt{\overline{Y_2^2}}$ &$\sqrt{\overline{Y^2}}$ & $|Y_1|_{\max}$ & $|Y_2|_{\max}$ & $|Y|_{\max}$ \\
\hline
Revised   & 10--20\% & $0.327\pm0.014$ & $0.349\pm0.019$  & $3.575\pm0.067$  & $1.665$ & $2.203$ & $8.060$ \\
Erlang    & 20--30\% & $0.368\pm0.019$ & $0.341\pm0.018$  & $3.522\pm0.065$  & $2.373$ & $2.284$ & $7.918$ \\
          & 30--40\% & $0.368\pm0.018$ & $0.295\pm0.015$  & $3.656\pm0.066$  & $2.088$ & $1.817$ & $8.015$ \\
          & 40--50\% & $0.357\pm0.016$ & $0.339\pm0.016$  & $3.582\pm0.068$  & $1.985$ & $1.803$ & $7.996$ \\
\hline
Unrevised & 10--20\% & $0.306\pm0.013$ & $0.357\pm0.020$  & $3.569\pm0.067$  & $1.458$ & $2.523$ & $8.045$ \\
Erlang    & 20--30\% & $0.338\pm0.016$ & $0.353\pm0.021$  & $3.509\pm0.065$  & $1.997$ & $2.928$ & $7.866$ \\
          & 30--40\% & $0.351\pm0.018$ & $0.301\pm0.015$  & $3.637\pm0.066$  & $2.129$ & $1.876$ & $8.033$ \\
          & 40--50\% & $0.331\pm0.015$ & $0.349\pm0.017$  & $3.573\pm0.068$  & $1.985$ & $2.425$ & $7.996$ \\
\hline
\end{tabular}%
\end{center}}
\end{table*}

From Figs. 1--4 and Tables 1 and 2, one can see that the effect of
anisotropic emission in the transverse plane on the $p_T$ spectra
is not obvious, though $v_2$ is large in the considered four
centrality classes. With increasing the centrality percentage from
0\% to 60\% or 80\%, the parameters $p_0$, $n$, and $k_1$
increase, which reflects that the strength and fraction of the
hard process increase from the central to peripheral collisions
due to the decreasing participant region in the interaction system
and secondary cascade collisions in the soft process. At the same
time, the parameter $\langle p_{Ti} \rangle$ decreases or does not
change approximately from the central to peripheral collisions,
which reflects the less or nearly invariant energy deposition with
increasing the centrality percentage due to the limiting secondary
cascade collisions. It is natural or may be coincidental that the
value of $m$ is equal to the number of quarks in identified
particles. That is why $m=2$, 2, and 3 correspond to the
productions of $\pi^++\pi^-$, $K^++K^-$, and $p+\bar p$,
respectively.

From central to peripheral collisions, the parameters $a_x$ and
$|b_x|$ increase, the parameter $a_y$ is fixed to 1 due to our
requirement, and the parameter $|b_y|$ increases or does not
change approximately. These tendencies reflect that the source has
a larger expansion along the $Ox$ axis in peripheral collisions
than that in central collisions. The source has also a larger
movement along the $Ox$ axis or in the planes of the first and
third quadrants in peripheral collisions than that in central
collisions. The larger expansion and movement of the source in
peripheral collisions are resulted from the larger asymmetry in
geometry and mechanics. In addition, less participant region and
secondary cascade collisions in peripheral collisions can reduce
the interactions between or among different sources, which can
also reduce the probability for isotropy. As a result, peripheral
collisions present a more obvious anisotropic spectrum.

Although we have used the Erlang distribution to describe the
$p_T$ spectrum produced in the soft process, many other functions
can be used in the fit process, too. These functions include, but
are not limited to, the two-component standard (Boltzmann,
Fermi-Dirac, and Bose-Einstein) distribution [53], the Tsallis
distribution with different forms [53, 54], the distribution of
blast-wave model with different statistics [55, 56], etc. To
revise these functions for describing $v_2(p_T)$ distribution, we
can perform the same or similar treatment as what we do for the
Erlang distribution. In particular, the Tsallis distribution can
fit the sum of two- and even three-component standard distribution
[57, 58], which reveals the advantage of the Tsallis distribution.
However, the physics essence of the Tsallis distribution is needed
to undergo more studies. The relation between the temperatures
obtained from the standard distribution and the Tsallis
distribution is also of interest [57, 58].

From Fig. 5 and Table 3 one can see that, the central region or
source contributes a wide and major $\eta$ spectrum, which is
resulted from the Landau hydrodynamic model [8--19]. The backward
(forward) region or source contributes a narrow and minor $\eta$
spectrum, which is a revision for the Landau hydrodynamic model.
The distribution widths contributed by the central and backward
(forward) sources in central collisions are less than or equal to
those in peripheral collisions due to the stronger stopping power
of nucleus in central collisions or the effect of leading
nucleons. The $\eta$ shift of the backward (forward) source in
central collisions is less than that in peripheral collisions due
to the stronger stopping power. The stronger stopping power
corresponds to the weaker penetrating power and narrower $\eta$
spectrum. Comparing with peripheral collisions, more resonances
are produced in leading nucleons in the backward (forward) source
in central collisions due to more multiple scatterings undergone
by the leading nucleons, though more particles can be produced in
central collisions due to more energy depositions. As a
competitive result, it is possible that the contribution ratio of
the backward (forward) source in central collisions is greater
than or equal to that in peripheral collisions.

It should be noted that although we have used a three-Gaussian
function to describe the $\eta$ spectrum in Fig. 5, it is of
course not a sole and necessary choice. In fact, different
pictures and functions are used in different models to obtain the
same or similar results. Even different functions are used in the
same or similar hydrodynamic model [13--19]. In particular, in
refs. [13, 16] where no leading nucleons or stopping effects are
included while the distributions are well described. In refs.
[14--16], the participant dissipating energy picture is used and
the distributions are well described, too. However, in refs.
[17--19], two sources of leading particles are included in the
distributions. One of them contributes a Gaussian function in the
backward region and another one contributes a Gaussian function in
the forward region.

Moreover, although we study Pb-Pb collisions in the present work,
the same or similar method can be used in studies of proton-proton
and proton-nucleus collisions (with high multiplicity) due to
similarity of experimental spectra [21, 52, 59]. The similar
experimental spectrum for different types of collisions at high
energy reveals some universality in hadroproduction process, as it
is argued in refs. [14--16, 60, 61]. The universality in
hadroproduction process appears in different quantities measured
[62] in different types of collisions and/or at different
energies. These quantities include, but are not limited to, mean
multiplicity, rapidity or pseudorapidity density, multiplicity or
transverse momentum distributions.

Combining Figs. 6--8 and Tables 4--6, one can see that the effect
of anisotropic emission in the transverse plane on the event
patterns is not obvious, though $v_2$ is large in the considered
four centrality classes. In the considered four centrality
classes, the root-mean-squares and maximums for velocity,
momentum, and rapidity components do not depend on the centrality
obviously. Because of the anisotropic emission in the transverse
plane, we obtain a less root-mean-square in the $y$ component than
that in the $x$ component. In the case of considering isotropic
emission in the transverse plane, we can obtain equivalent
root-mean-squares in both the $y$ and $x$ components.

It is expected that the event patterns displayed by the scatter
plots of different particles produced in different centralities at
different energies have some similarities or differences. In
particular, the event patterns displayed by the scatter plots of
charged particles in the three-dimensional velocity space are
rough sphericity (or fat ellipsoid along $Oz$ axis) with high
density close to $\beta_z=1$ [22]. This rough sphericity does not
depend obviously on the centrality at dozens of GeV and above. The
event patterns displayed by the scatter plots of $Z$ bosons or top
and anti-top systems are rough cylinder [21, 23]. This rough
cylinder does not depend obviously on the centrality in the energy
range mentioned above.
\\

{\section{Conclusions}}

We summarize here our main observations and conclusions.

a) We have used the hybrid model to fit the $p_T$ spectra,
dependence of $v_2$ on $p_T$, and $\eta$ spectra of charged
particles produced in Pb-Pb collisions at $\sqrt{s_{NN}}=2.76$
TeV. At the same time, the $p_T$ spectra and dependence of $v_2$
on $p_T$ for identified particles are fitted by the model. The
model results are approximately in agreement with the experimental
data of the ALICE Collaboration. All parameter values and event
patterns extracted from the fits reflect the properties of
interaction system at the stage of kinetic freeze-out, but not
those at the stage of chemical freeze-out, due to these values and
patterns being extracted from the $p_T$ and $\eta$ spectra as well
as dependence of $v_2$ on $p_T$.

b) From central to peripheral collisions, the strength and
fraction of the hard process increase and the energy deposition
decreases or is nearly invariant. The emission source in
peripheral collisions has a larger expansion along the $Ox$ axis,
and has also a larger movement along the $Ox$ axis or in the
planes of the first and third quadrants. Finally, peripheral
collisions present a more obvious anisotropic spectrum. Although
$v_2$ is large in some cases, the effect of anisotropic emission
in the transverse plane on the $p_T$ spectra is not obvious.

c) In the fit to $\eta$ spectra by the three-Gaussian function,
the central source contributes a wide and major $\eta$ spectrum,
and the backward (forward) source contributes a narrow and minor
$\eta$ spectrum. In central collisions, the distributions
contributed by the central and backward (forward) sources have a
less or invariant width and the $\eta$ shift of the backward
(forward) source is less. Meanwhile, more resonances are produced
in leading nucleons in the backward (forward) source, though more
particles can be produced. As a competitive result, the
contribution ratio of the backward (forward) source in central
collisions is possibly greater than or equal to that in peripheral
collisions.

d) The event patterns displayed by the scatter plots of charged
particles in the three-dimensional velocity space are rough
sphericity (or fat ellipsoid along $Oz$ axis) with high density
close to $\beta_z=1$. The event patterns in the three-dimensional
momentum space are rough cylinder with some removed particles from
the profile. The event patterns in the three-dimensional rapidity
space are rough cylinder with a high peak and long tail at the top
and bottom respectively. These observations do not depend
obviously on the centrality. The effect of anisotropic emission in
the transverse plane on the event patterns is not obvious, though
$v_2$ is large in the considered four centrality classes.
\\

{\bf Conflict of Interests}

The authors declare that there is no conflict of interests
regarding the publication of this paper.
\\

{\bf Acknowledgments}

Comments on the paper from the referee and communications from
Edward K. G. Sarkisyan are highly acknowledged. This work was
supported by the National Natural Science Foundation of China
under Grant No. 11575103, the Shanxi Provincial Natural Science
Foundation under Grant No. 201701D121005, and the Fund for Shanxi
``1331 Project" Key Subjects Construction.
\\

\end{multicols}
\end{document}